\documentclass{article}
\usepackage[english]{babel}
\usepackage[latin1]{inputenc}
\usepackage{graphicx}
\usepackage{pstricks}
\usepackage{dcolumn}
\usepackage{bm}
\usepackage{geometry}
\usepackage{amssymb,amsmath}
\title{Turbulent spot growth in plane Couette flow: statistical study and formation of spanwise vorticity}
\date{01/14/2014}

\author{Joran Rolland \footnote{LadHyX, UMR CNRS 7646, \'Ecole Polytechnique, 91128 Palaiseau, France ;
 INLN, UMR CNRS 7335, UNSA, 1361 route des lucioles, 06560 Valbonne, France}
\footnote{joran.rolland@ladhyx.polytechnique.fr}}
\begin{document}

\maketitle
\begin{center}
{\footnotesize Keywords : transition, wall-bounded turbulence, shear flows instabilities, 

}
\end{center}
\begin{abstract}
This article presents direct numerical simulations of the growth of turbulent spots in the transitional regime of plane Couette flow. A quantitative description of the growth process and of the detail of the quadrupolar flow around the spot is given. Focus is put on formation of spanwise vorticity in the velocity streaks that resembles a secondary shear instability. The main features of the instability (coherence lengths, advection velocity) are studied in the context of the turbulent spot, below and above the threshold Reynolds number of appearance of the oblique turbulent bands of plane Couette flow.
\end{abstract}

\section{Introduction}

   In shear flows such as plane Couette flow (Fig.~\ref{sk}), the discontinuous nature of the transition to turbulence allows the spatial coexistence of laminar and turbulent flow. The spot stage corresponds to the expansion in two directions of
turbulence, triggered by a strong enough local perturbation of the laminar flow. Spot expansion is possible provided the Reynolds number
is over $R_{\rm g}\simeq 325$, the Reynolds number below which turbulence
cannot exist \cite{prigent,RM}. The spot expansion leads to the oblique band regime \cite{prigent,RM}
provided $R<R_{\rm t}\simeq 415$ \cite{prigent} or
completely invades the domain if $R>R_{\rm t}$. Spot growth in plane
Couette flow has been studied numerically in a large range of Reynolds number \cite{lJ}. Further studies showed the quadrupolar flow around the spot and the Reynolds stress generating the spot \cite{lmpof} or the labyrinthine growth in very large domains \cite{DSH}.

  This note is interested in a departure from the streaks \& vortices flow
that has been reported in the band regime of plane Couette
flow \cite{etc}. This secondary instability of the velocity streaks is very similar to what is found in Hagen--Poiseuille pipe flow \cite{SK,DWK}. This  departure resembles a secondary Kelvin--Helmholtz instability of the streamwise velocity which forms rolls characterised by spanwise vorticity.
The preliminary study of the instability showed that linear analysis rendered the phenomenon well. Like other studies \cite{PRL}, this note will stress on the advection by the large scale flow around the band or the spot. Indeed, the effect of an unsteady large scale flow on the advection of the rolls, as well as its behaviour for $R>R_{\rm t}$ are unknown. Spot growth is an
interesting system in that case, since it is the only regime where
laminar and turbulent flow coexist at large Reynolds numbers, and where a net large scale
flow can be found.

  In order to investigate these questions, this article is organised as follow:
the numerical procedure and the system are described in a first
section (\S~\ref{meth}). Two examples of growth and reorganisation
of the spot, corresponding to two Reynolds numbers above and below $R_{\rm t}$, are studied quantitatively. The instability
is then described, first qualitatively on snapshots, then quantitatively
using procedures that will be studied systematically in a future article \cite{isp}
(\S~\ref{inst}). Results are discussed is section~\ref{D}.

\section{Method \label{meth}}

\begin{figure}\centerline{\includegraphics[height=3cm]{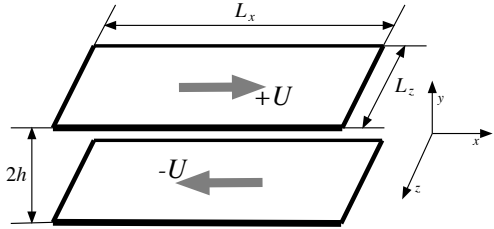}}
\caption{Sketch of plane Couette flow}
\label{sk}
\end{figure}

  The system is formally described by the incompressible Navier--Stokes equations
with plane Couette base flow and in plane periodic boundary
conditions (Fig.~\ref{sk}). The directions $\mathbf{e}_x$, $\mathbf{e}_y$, $\mathbf{e}_z$ are respectively the streamwise, wall-normal and spanwise directions. The
Reynolds number $R=hU/\nu$ is the
control parameter, with $h$ the half gap, $U$ the
velocity of the top plate, and $\nu$ the kinematic viscosity. The velocities, lengths and times are made dimensionless respectively by $U$, $h$ and $h/U$. The departure $\mathbf{v}$ from the plane
Couette baseflow $y \mathbf{e}_y$ is the variable of interest. The square of the norm of the departure $\parallel\mathbf{v}\parallel^2$ is termed the energy.

Two Reynolds numbers, $R=370<R_{\rm t}$ and
$R=500>R_{\rm t}$ are considered here. The size of the periodical
domain is $L_x\times L_z=110\times 80$. It is adapted to
accommodate one band at $R=370$ at this resolution \cite{RM}.
The Navier--Stokes equations are integrated in time using the {\sc ChannelFlow} code
by John Gibson \cite{gibs}. The same resolution ($N_y=27$ Chebyshev modes and $4$ dealiased Fourier modes per unit length) is taken. It is
sufficient for a good rendering of the bands around $R_{\rm
t}$ \cite{MR}.

  A simple initial condition is used to create turbulent spots.
A small square of size $l_x\times l_z=8\times 8$ containing the
average turbulent profile is at the center of the domain, in the
middle of otherwise laminar flow. Two different initial conditions are created at $R=370$ using two realisations of a spatially white additive noise.

\section{Statistical description of the flow \label{stat}}

\subsection{Procedures}

   In order to fix the ideas on the spatial organisation of turbulence, we first consider colour levels of the
energy (see figure~\ref{f1_bis}, at $R=370<R_{\rm t}$ (a,b), $T=200$ after the beginning of the simulation and at $R=500>R_{\rm t}$
(c,d), $T=150$ after the beginning of the simulation). They show that for both
Reynolds numbers, turbulence in the upper part of the flow is shifted with respect to
turbulence in the lower part. This feature is found for bands as
well \cite{RM} and leads to the so called overhanging regions \cite{coles66}. Given the organisation of the flow \cite{RM,isp}
(Fig.~\ref{f1_bis}), we define the laminar, intermediate
and turbulent zones. The laminar zone corresponds to no departure from
the the laminar base flow. The intermediate, or overhanging, zone corresponds to
turbulence on top or below laminar flow \cite{coles66}. The turbulent
zone corresponds to turbulence over the whole gap.

  In order to provide quantitative definitions of these zones, we use a coarse
graining and thresholding procedure. It has been introduced in a former article \cite{RM}, and discriminates between laminar and
turbulent locations in the flow. The flow is divided in cells of size $l_x\times l_y \times l_z=2\times 1\times 2$, $y>0$ or $y<0$, in which the kinetic energy $\parallel \mathbf{v}\parallel^2$ is averaged. If it is above $c=0.025$, the cell is turbulent, otherwise, it is laminar. This allows one to define the state of the flow at position $(x,z)$: if two laminar cells are on top of each other, we are in the laminar zone, if two turbulent cells are on top of each other, we are in the turbulent zone. If a laminar cell is on top of a turbulent one (or \emph{vice versa}) we are in the intermediate or overhanging zone.

This discrimination procedure is used to compute the
turbulent fraction $f$, which measures the volume occupied by
turbulence, \emph{i.e.} the fraction of turbulent cells. The discrimination procedure is used to conditionally average $\parallel\mathbf{v}\parallel^2$ in the turbulent cells. This yields $e_t$, the
turbulent energy. It measures the intensity of turbulent
inside the bands or spots. The square of the norm $\parallel\mathbf{v}\parallel^2$ (kinetic energy) averaged over the
whole flow is simply denominated as average energy $e$. This
discrimination procedure allows one to compute the area $f_f$ projected on the planes $y=\pm1$
occupied by the intermediate zone between turbulent and
pseudo-laminar flow \cite{isp}. This is the number of $y<0$ turbulent cells facing a $y>0$ laminar one plus that of $y>0$ turbulent cells facing a $y<0$ laminar one, divided by half the number of cells. Note that $f$ and $f_f$ represent respectively a volume and a surface which is part of the boundary of said volume.

The order parameter $m_\pm$ is defined as the wall normal average of the square of the fundamental mode of the band in the Fourier
transform of the streamwise velocity field \cite{RM,holes}:
\begin{equation}m_\pm^2=\frac{1}{2}\int_{-1}^1|\hat{v}_x|^2(k_x,\pm k_z) {\rm d}y\,.\end{equation}
It measures the
intensity of the modulation of turbulence in the flow. The $\pm$ sign corresponds to the $k_x,\pm k_z$, \emph{i.e.} the two orientations of the band. In our case, the modes
$k_x=1$, $k_z=\pm1$ are the most relevant. If $m_\pm$ is larger than $m_\mp$, this means that one orientation dominates the other. If they are nearly equal, this means that both orientations have the same weight in the spot.

These
procedures are validated at low and normal resolution \cite{RM}. Lower case letters ($f$ \emph{etc}.) denote instantaneous values, while capitalised ones ($F$ \emph{etc}.) denote time averages performed when the flow has reached a steady state.

\subsection{The growth process}

\begin{figure}
\centerline{{\large
\textbf{(a)}\hspace{1mm}\includegraphics[width=3cm,clip]{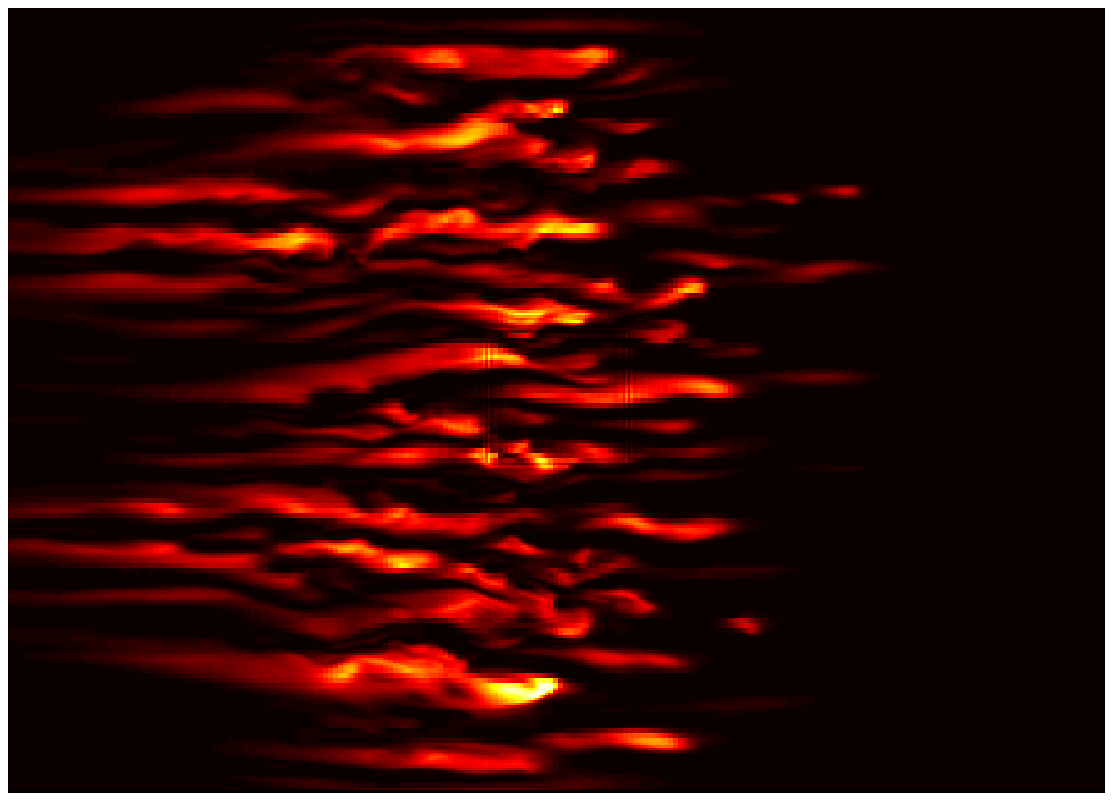}\hspace{0.1cm}\textbf{(b)}\hspace{1mm}}\includegraphics[width=3cm,clip]{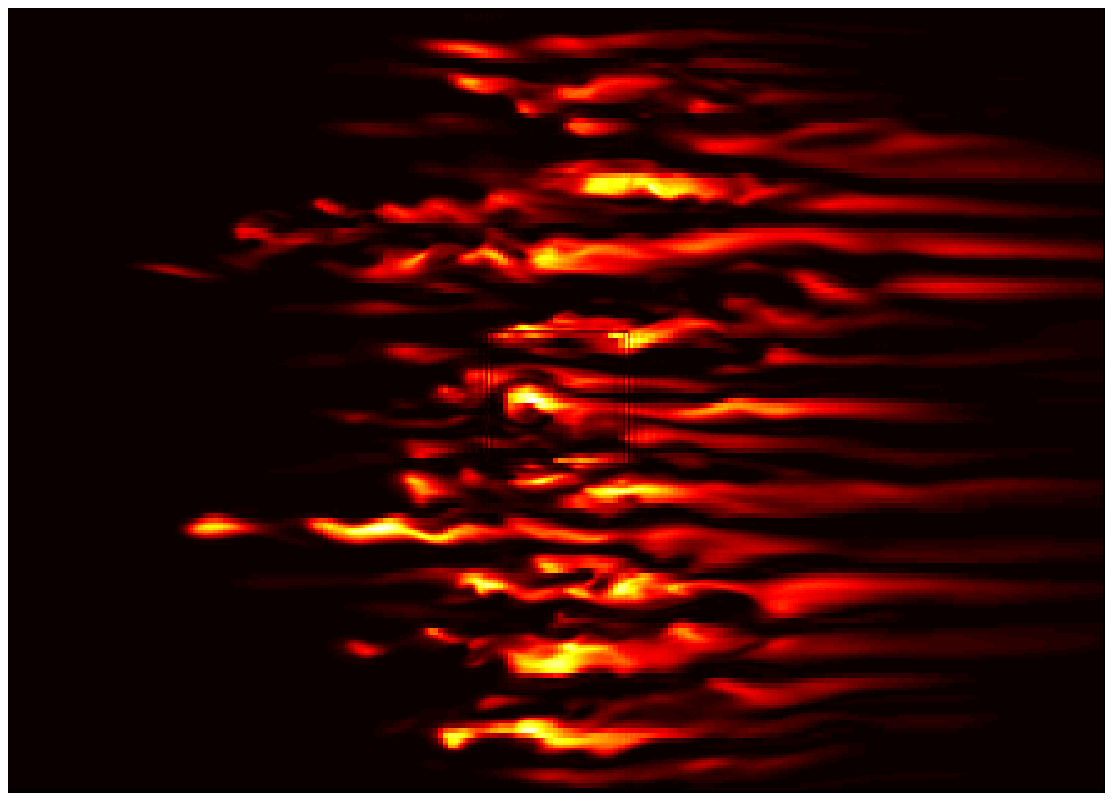}
 {\large
\textbf{(c)}\hspace{1mm}\includegraphics[width=3cm,clip]{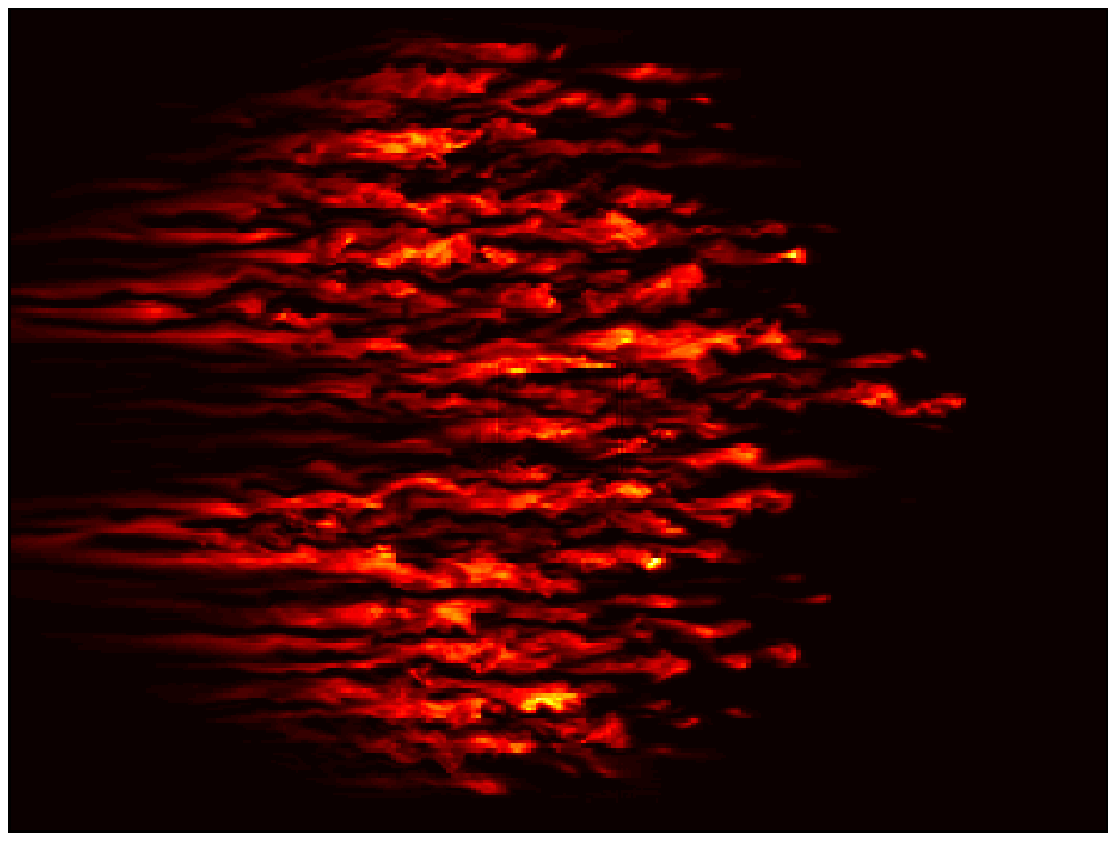}\hspace{0.1cm}\hspace{0.1cm}\textbf{(d)}\hspace{1mm}}\includegraphics[width=3cm,clip]{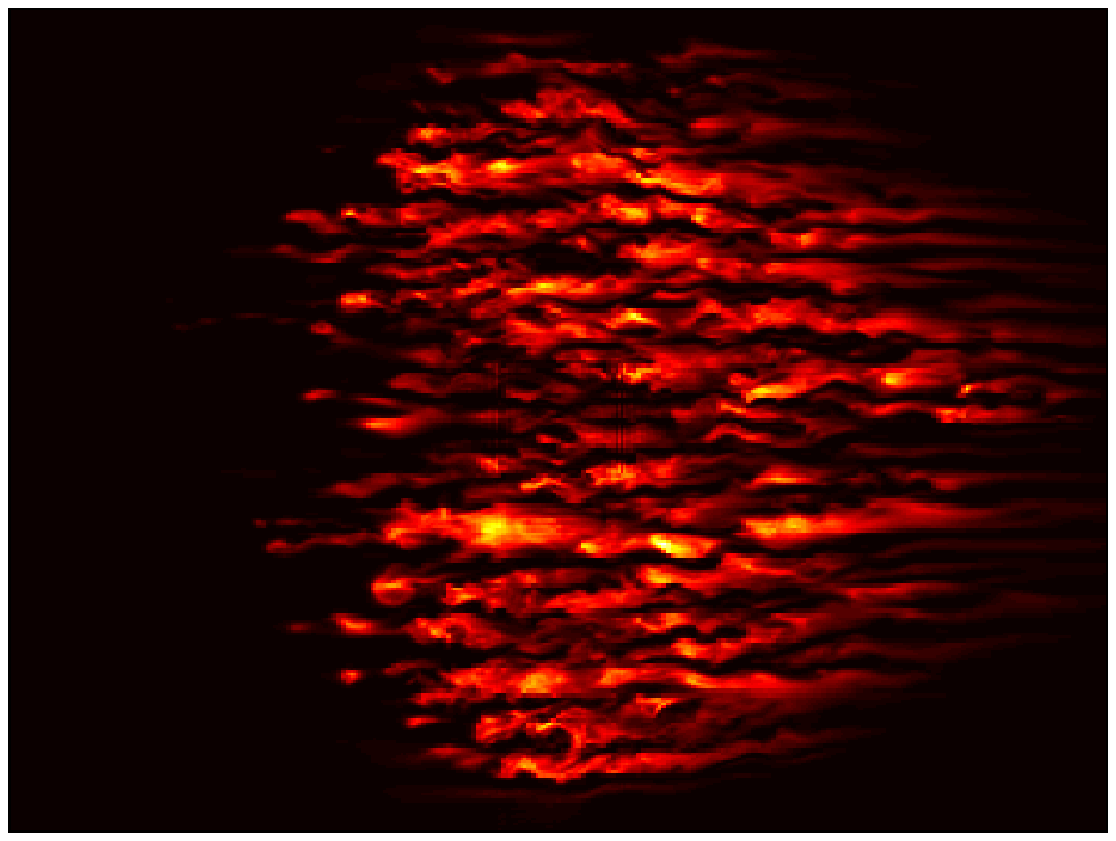}\includegraphics[height=2.5cm,clip]{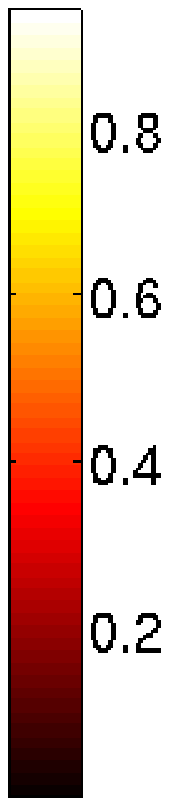}}
\caption{Examples of spots at $R=370$ (a, b) and $R=500$ (c, d)
before the perturbation reaches the boundaries of the domain, colour
plot of $\bf{v}^2$ in $y=-0.6$ (a, c) and $y=0.6$ (b, d) plane in a
periodic domain $L_x=110$, $L_z=80$} \label{f1_bis}
\end{figure}

\begin{figure}
\centerline{\includegraphics[width=6.5cm,clip]{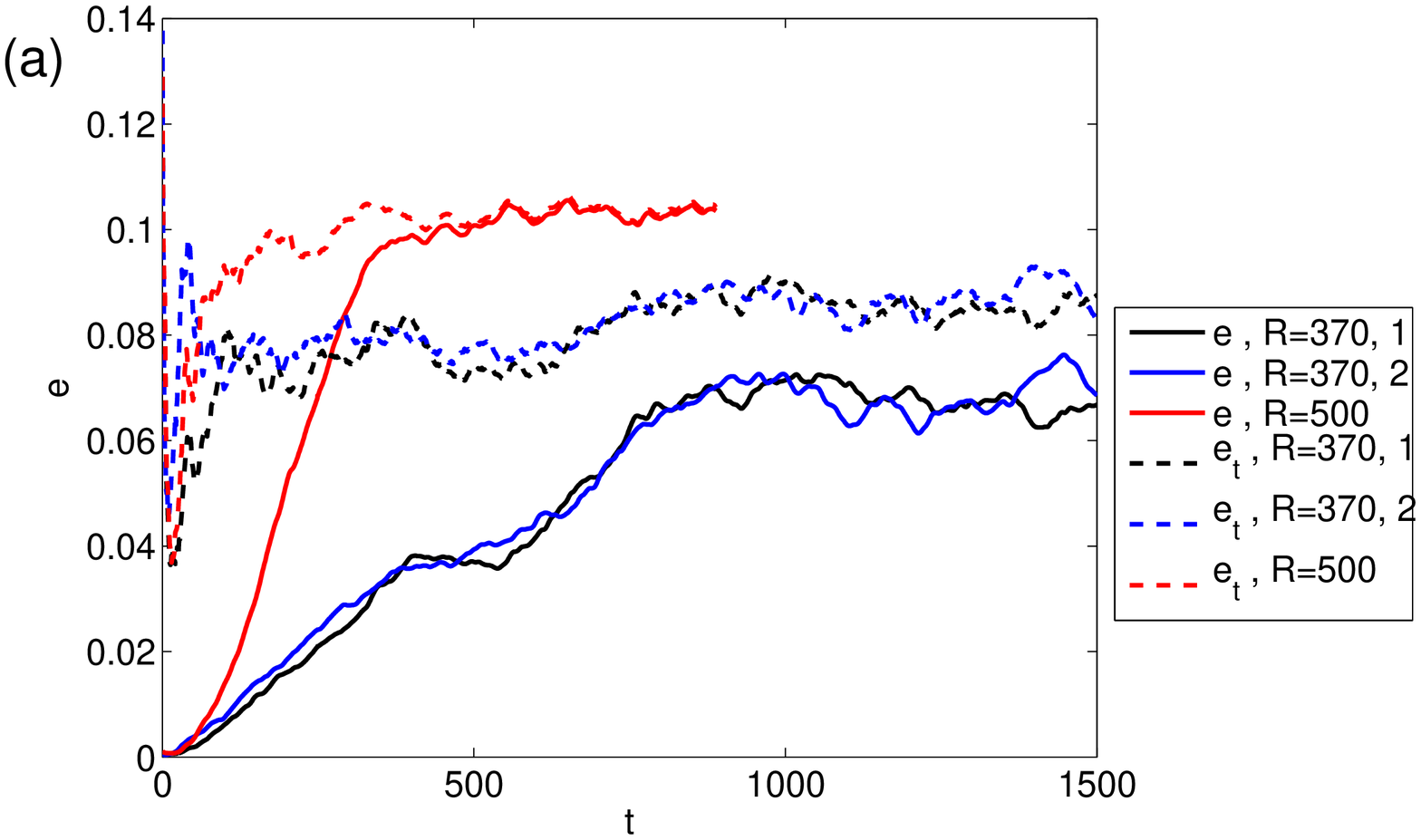}
\includegraphics[width=6.5cm,clip]{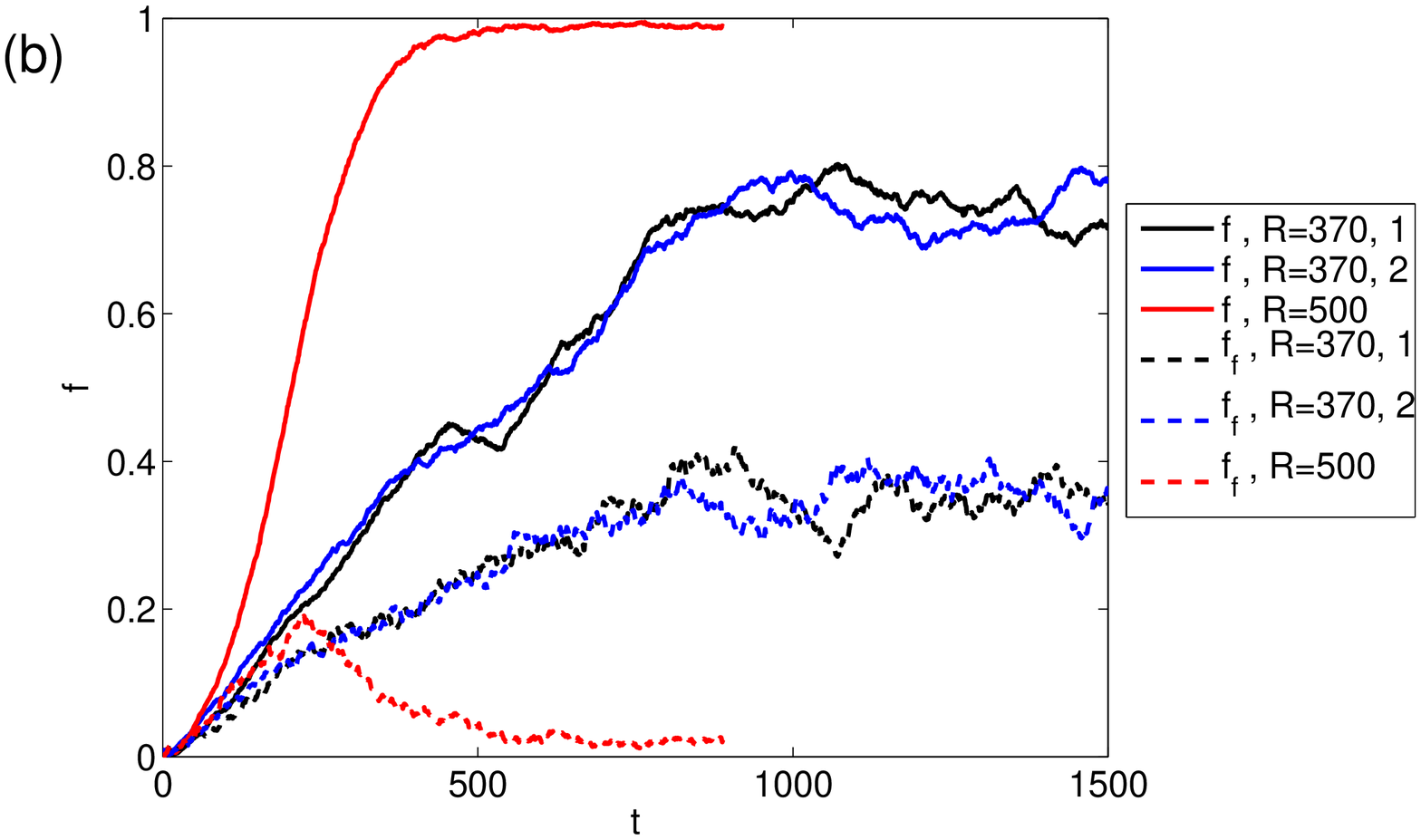}
\includegraphics[width=6.5cm,clip]{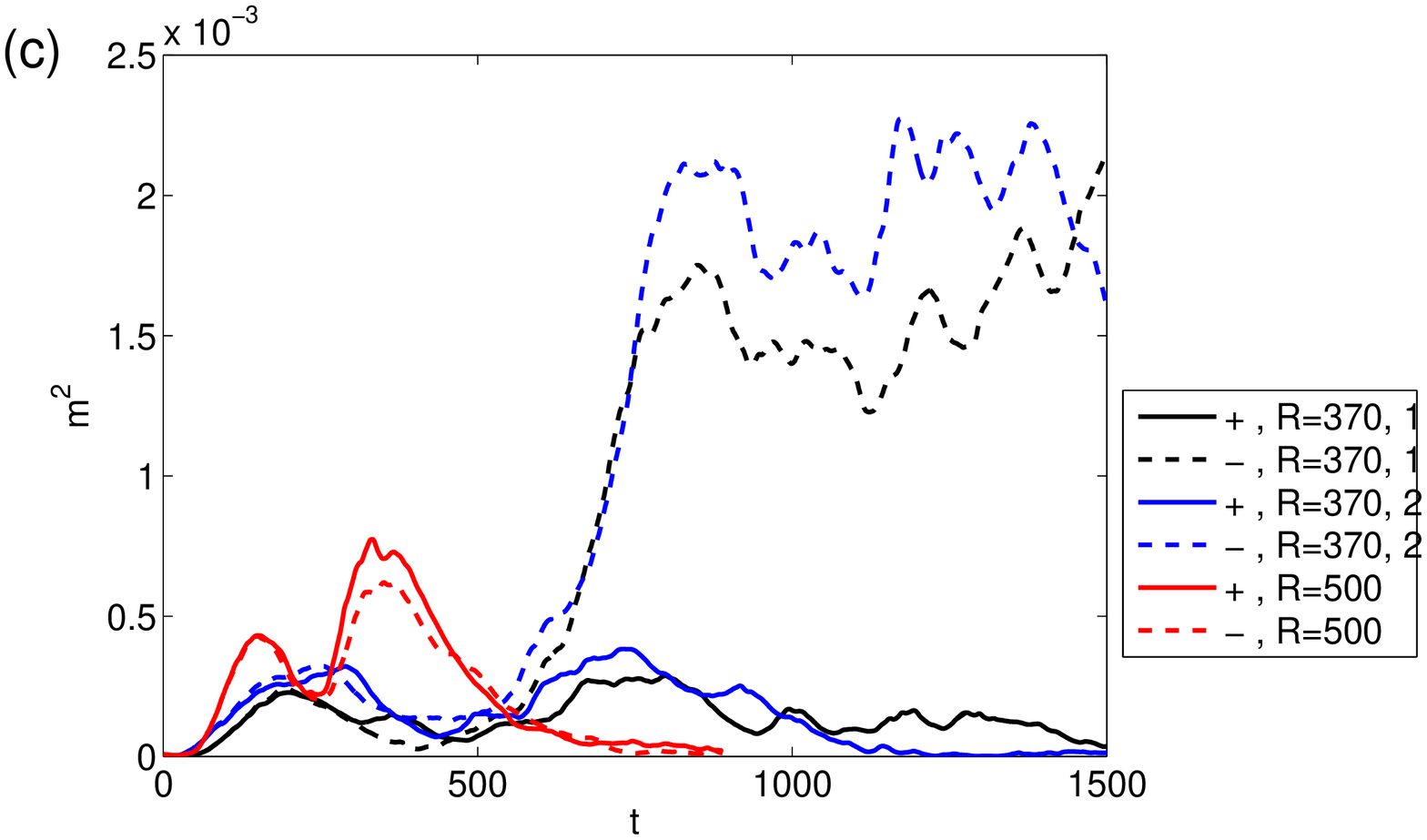}}
\caption{At $R=370$ and $R=500$, Time series of (a): energy and turbulent energy, (b) turbulent fraction and intermediate area, (c) : Fourier mode (order parameter)}
\label{timeserie}
\end{figure}

  The average quantities are used to monitor the growth of the spot (Fig.~\ref{timeserie}). The average
and turbulent energies are displayed in figure~\ref{timeserie} (a).
At both Reynolds numbers, the turbulent energy reaches the
neighbourhood of its average value after a short time ($T\lesssim
50$).  This simply indicates that the flow quickly goes
from the somewhat artificial initial condition to low Reynolds
number turbulent flow. The average value $E_t$ increases with increasing Reynolds
number \cite{RM}. Local turbulence is more intense as $R$ is
increased. The flow remains as such inside the band : this will be seen in profiles conditionally
averaged in space as well.  At
$R=500$ the average energy grows monotonically with time and then
reaches a plateau value ($T\sim$500). It increases faster than at
$R=370$. This is consistent with earlier findings \cite{lJ}. Both experiments at $R=370$ display a very similar growth. However, the average energy has a much clearer plateau ($T\sim$400) then a small decrease
($T\sim$600) in one of the two numerical experiments. This may be a consequence of the random nature of the growth \cite{Yms}. It mainly increases before reaching the equilibrium value $E$ at $T\sim 800$. This last stage
corresponds to the reorganisation of the spot in a band after it
feels the boundary conditions.

The behaviour of the turbulent fraction $f$ is the same as
that of $e$ (Fig.~\ref{timeserie} (b)). Indeed,
one has $e\simeq e_t\times f$ \cite{RM}. At $R=500>R_{\rm t}$, the
equilibrium corresponds to a featureless turbulent flow, with
$F\lesssim 1$. At $R=370<R_{\rm t}$, the equilibrium regime (in the
sense of fixed turbulent fraction) corresponds to modulated
turbulence, with $F\simeq 0.7$. This is consistent with the thorough study of the band configuration \cite{RM}. The size of the intermediate area $f_f$ increases then decreases
at $R=500$: the intermediate zones between laminar and turbulent
form, but then disappear because the spot invades the whole flow.
Meanwhile, at $R=370$ $f_f$ reaches a plateau value after the
reorganisation. The last stage is only seen at $R=370$ ($T\gtrsim 600$). It
corresponds to the reorganisation of the modulation of turbulence
into one band.

   Time series of the square of the order parameter $m_\pm^2$ are recorded (Fig.~\ref{timeserie} (c)). In the first stages
of the spot growth, the Fourier modes corresponding to both
orientation are approximately equal, this is the consequence of the symmetry with respect to the $z=40$ plane of the initial condition. The small difference is the consequence of
the small spatial asymmetries of the initial condition and the locally chaotic evolution. The order
parameters $m_{\pm}^2$ grow for both Reynolds number due to the
appearance of an oblique structure in the flow, \emph{i.e.} the
so-called arrow shaped spot \cite{lJ}. This stage takes place up to $T\sim 350$ at $R=500$. The order parameters $m_\pm$ then decreases to
small non zero value. There is but a small trace of the modulation
of turbulence, caused by turbulent fluctuations \cite{RM}. At
$R=370$, both order parameters are nearly equal up to $T\simeq 600$. The flow then loses symmetry, as shown by the increasing difference between $m_+$ and $m_-$, and there is a competition between
orientations. It settles quickly in that case ($T\sim 1000$): the competition is
eventually won by the $-$ orientation. Note that the beginning of the symmetry breaking occurs at the same time of the plateau of the kinetic energy.

  The elimination of small defects of the band takes place in the range $800\lesssim T \lesssim 2000$. The competition and elimination of defects is precisely modeled
by a Landau--Langevin model \cite{RM2}. Note that during the symmetrical growth stage, the diamond shaped spot seems to be composed of four ``half bands'': two of each orientation in each half gap, shifted by the laminar baseflow (Fig.~\ref{f1_bis}). This means that the symmetry breaking leading to a purely band shaped spot requires some reorganisation.

  The overall evolution of a system that contains one band at statistical equilibrium is the same as that of a system that can contain a
large number of bands \cite{DSH}. In both cases, one can find the
invasion/reorganisation phases. The flow eliminates the disclinations and most of the laminar holes and turbulent bulges to fit the boundary conditions. However, the reorganisation stage can be
longer due to the larger number of defects that have to be
eliminated \cite{DSH}.

\subsection{The large scale velocity field}

We endeavour to  improve the former descriptions of velocity profiles in the spot, that were mainly two-dimensional \cite{lJ,lmpof}, and discuss possible symmetries of the velocity field similar to those found in the bands \cite{BT07}. Besides, this will be of interest when we will examine the advection of rolls created by the secondary instability.

\begin{figure}\centerline{
\includegraphics[height=5cm,clip]{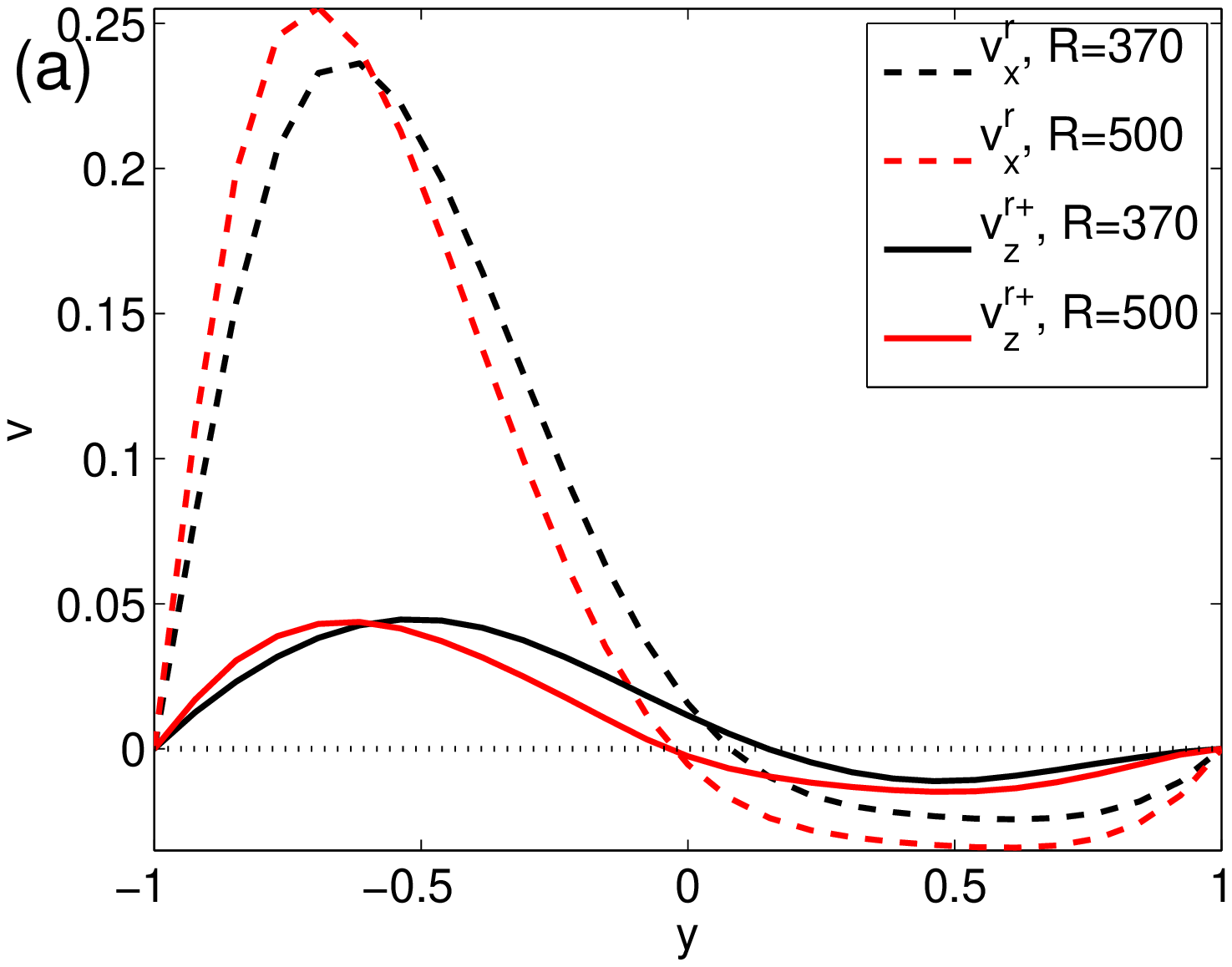}{\Large\textbf{(b)}\hspace{1mm}}\includegraphics[width=6cm,clip]{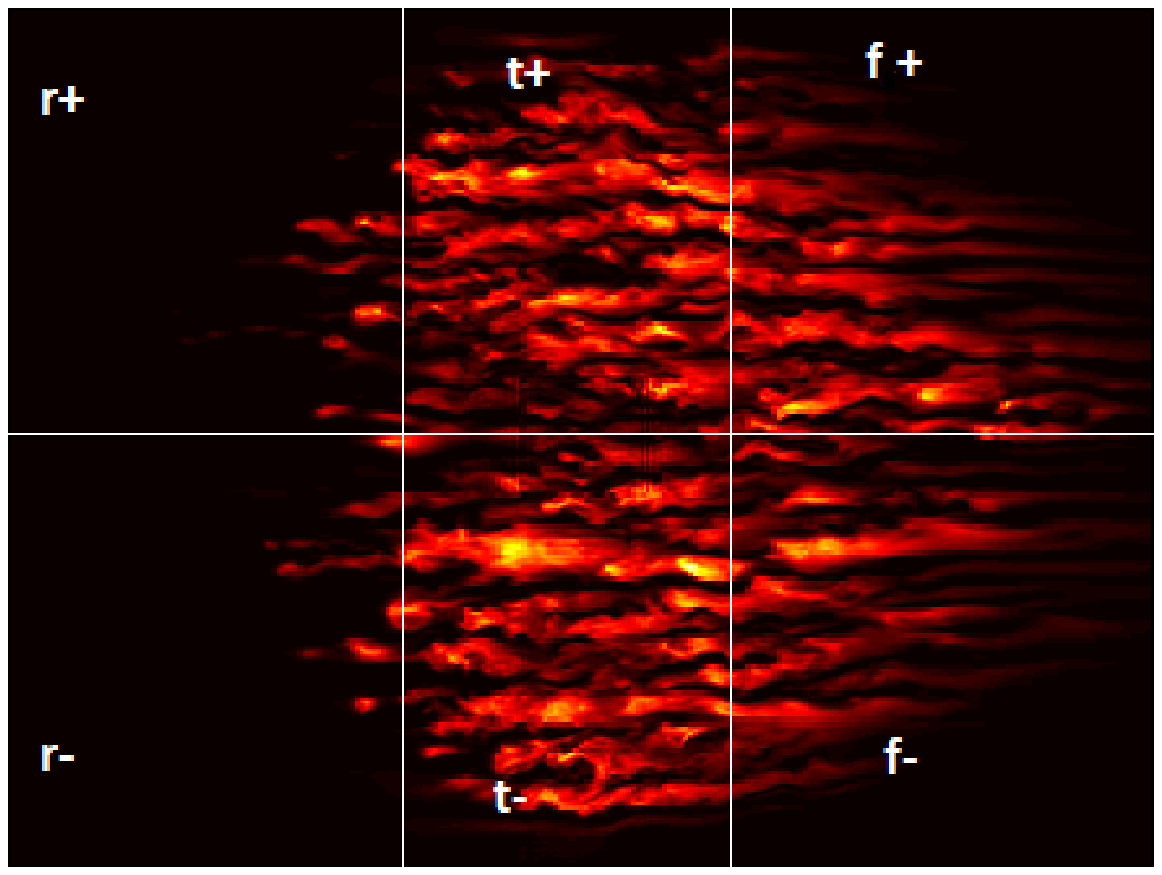}\includegraphics[height=5cm,clip]{echelle.eps}}
\caption{Streamwise and spanwise velocity profiles, for both Reynolds numbers,
conditionally averaged in the $r+$ intermediate zone of spots at Reynolds number
$R=370<R_{\rm t}$ and $R=500>R_{\rm t}$. (b) Colour levels of $\parallel\mathbf{v}\parallel^2$ at $y=0.62$ for $R=500$, on which is added the position of the quadrants f/t/r, +/-.} \label{average_spot}
\end{figure}

We use the laminar/intermediate/turbulent discrimination used in the former section to conditionally average velocity profiles. We distinguish the two intermediate zones: the one where laminar flow is on top of turbulent flow is called ``rear'' ($\rm r$) with respect to the direction $\mathbf{e}_x$ and the one where turbulent flow is on top of laminar flow is called ``front'' ($\rm f$). Due to the unsteady nature of the spots, a time average is only performed in the short
time window for which the intermediate zone exists
(Fig.~\ref{timeserie} (b)). This procedure is applied to both velocity fields $v_x$ and $v_z$ (Fig.~\ref{average_spot} (a)). Care is taken to average the spanwise profiles separately in the two parts of the domain delimited by
$z=40$ (plane of symmetry of the spot), so as not to cancel out the two antisymmetric contributions.

We describe the symmetries of the velocity field. The spanwise component $v_z^+$ is positive
for $z>40$ and $v_z^-$ is negative for $z<40$, \emph{i.e.}
$v_z^{+}\simeq -v_z^{-}$ (Fig.~\ref{average_spot} (b)). We find that $v_x^{\rm f}(y)=-v_x^{\rm
r}(-y)$, which corresponds to the $v\leftrightarrow -v$,
$y\leftrightarrow -y$ centro-symmetry \cite{BT07}. The S shaped turbulent profiles verify $v_x^{\rm
t}(y)=-v_x^{\rm t}(-y)$ and $v_x^{\rm t}(y)=v_x^{\rm f}(y)+v_x^{\rm
r}(y)$. One finds $v_z^{{\rm f},\pm}(y)=-v_z^{{\rm
r},\pm}(-y)$ and $v_z^{\rm t}(y)=v_z^{\rm f}(y)+v_z^{\rm r}(y)$.
This allows one to reconstruct the whole in plane and wall normal
dependence of the large scale flow with only two profiles (Fig.~\ref{average_spot} (a)). The quadrupolar flow around the early
spot, corresponding to the wall-normal average of this velocity field, is found \cite{lmpof}. The value of the wall normal average of such profiles is typically the half of the value of the maximum \cite{isp}. The shape of the profiles are unchanged whether $R$ is
smaller or larger than $R_{\rm t}$. As it is expected, as $R$ is increased, the amplitude of the velocity profiles is slightly more intense and the shear layers are smaller.

These data are in agreement with and complete
the profiles obtained numerically in the spots by Lundbladh
\& Johansson \cite{lJ} who display the total streamwise velocity field $v_x+y$, and very similar to what is found in the bands \cite{BT07}.

\section{The instability\label{inst}}

\subsection{Visualisation}

\begin{figure}
\centerline{\includegraphics[width=14cm,clip]{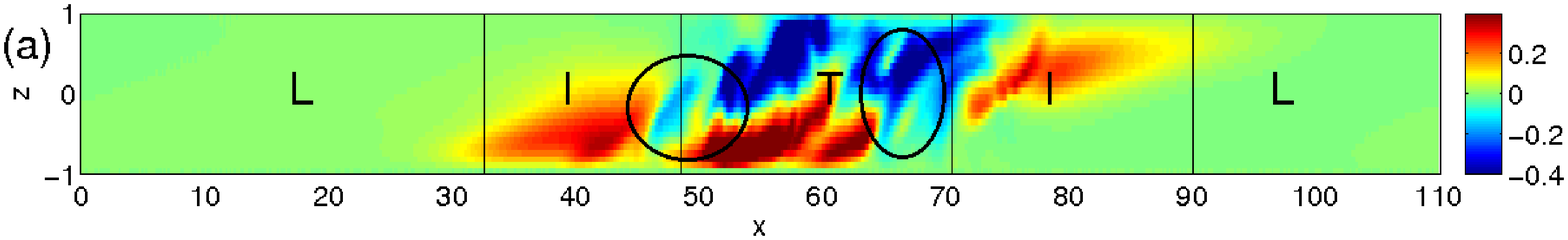}}
\centerline{\includegraphics[width=14cm,clip]{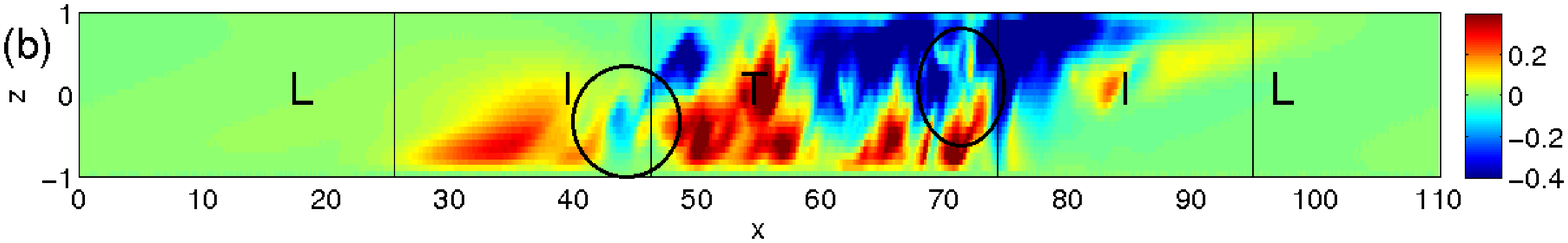}}
\caption{Two examples of colour levels of the streamwise velocity field showing the secondary instability in spots, at both Reynolds numbers. (a): $R=370<R_{\rm
t}$, (b): $R=500>R_{\rm t}$} \label{exspot}
\end{figure}

  The same approach as in the band case is taken \cite{etc}.
The instability is first identified in snapshots of the streamwise
velocity. Two examples are taken, one at $R=370$ and the other at
$R=500$. Colour
levels of the streamwise velocity field in $x-y$ planes are used
(Fig.~\ref{exspot}). Two $x-y$ planes are chosen at the early stage of
development of the spot. The velocity streaks baseflow \cite{etc,isp} and shift of turbulence in  upper
and lower part \cite{RM,etc,isp,BT07} can be seen. The laminar,
turbulent and intermediate zones are denoted by L, T and I.

  Examples of the instability developing in the
intermediate and turbulent zones can be seen in figure~\ref{exspot}. They are identified by the black circles.
The clearest examples can be seen in the intermediate zones. The
instability develops at $65\le x \le 70$ at $R=370$
(figure~\ref{exspot} (a)) and at $45\le x \le 50$ at $R=500$
(figure~\ref{exspot} (b)), the entanglement of velocity is typical
of a Kelvin--Helmholtz instability \cite{SK,DWK,isp}. Both cases are
very similar to what can be seen in the bands \cite{isp}. The
wavelengths are of order $5h$, although they appear to be smaller at
$R=500$.

  The turbulent area is perturbed by this instability as well. It can be seen at $x=60$ or $65$ at $R=370$,
  and at $x=65$ and $x=72$ at $R=500$.
Similarly to the band situation, the wavelength of the instability
is smaller.

\begin{figure}
\centerline{\includegraphics[width=14.5cm,clip]{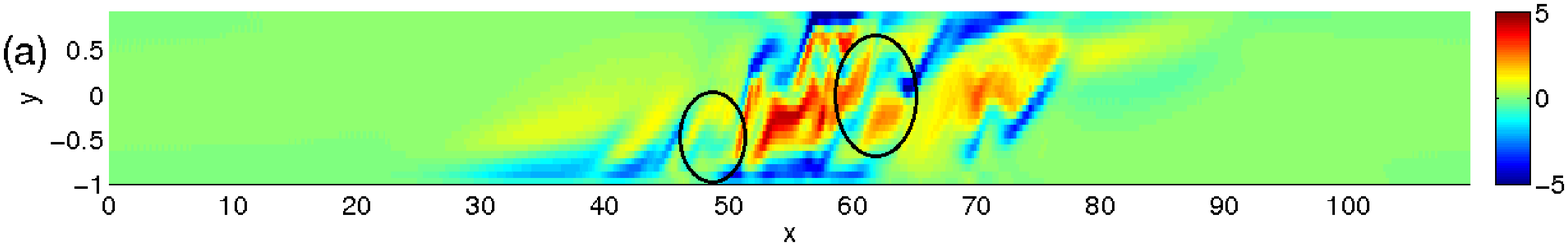}}
\centerline{\includegraphics[width=14.5cm,clip]{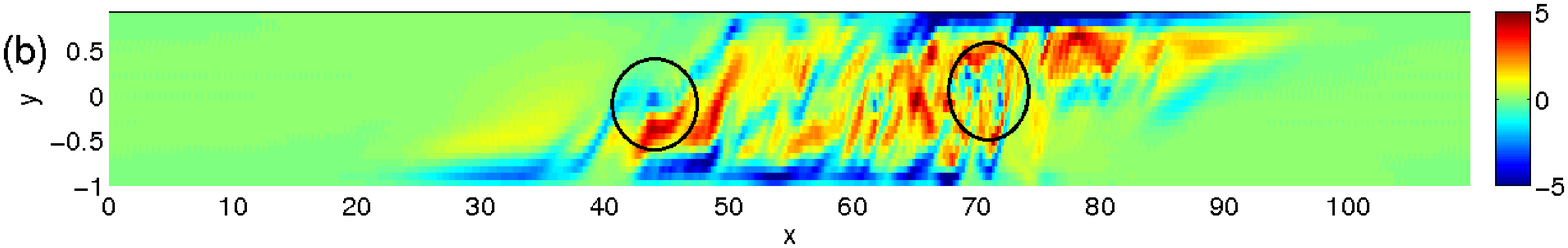}}
\caption{Spanwise vorticity fields corresponding to the streamwise velocity fields of figure~\ref{exspot}, (a): $R=370$, (b): $R=500$}
\label{exvort}
\end{figure}
  The detailed kinematic study of the case of the band shows that the secondary instability develops into rolls that are characterised by negative spanwise vorticy in the midgap \cite{isp}. This will provide a systematic and quantitative way to detect the instability beyond visualisations that are not always as clear in turbulent DNS as in the study of an isolated shear layer. A generic criterion is made more necessary by the fact that, unlike pipe flow, the perturbations move in the spanwise direction and are difficult to follow in snapshots of $x-y$ planes. The effect on vorticity appears when comparing the entangled streamwise velocity field (Fig.~\ref{exspot}) to the corresponding spanwise vorticity fields (Fig.~\ref{exvort}). The rolls stand out in the vorticity, because in the case of the velocity streaks, the shear $\partial_yv_x$ dominates in $\omega_z$. Starting from the average velocity profiles \cite{BT07}, one can easily see that velocity streaks lead to $\omega_z(y\simeq 0)>0$. Negative vorticity in the midplane is therefore expected of such rolls and in agreement with the formation of azimuthal vorticity found in pipe flow \cite{DWK,SK}. Note that this negative spanwise vorticity is different from the tongues describes by Schoppa \& Hussain in boundary layers \cite{schhu}, in that they are often disconnected from the wall. When they do reach out to the near wall, $\omega_z<0$ vorticity is stronger in the midgap.

\begin{figure}
\centerline{\textbf{(a)}\includegraphics[width=6cm]{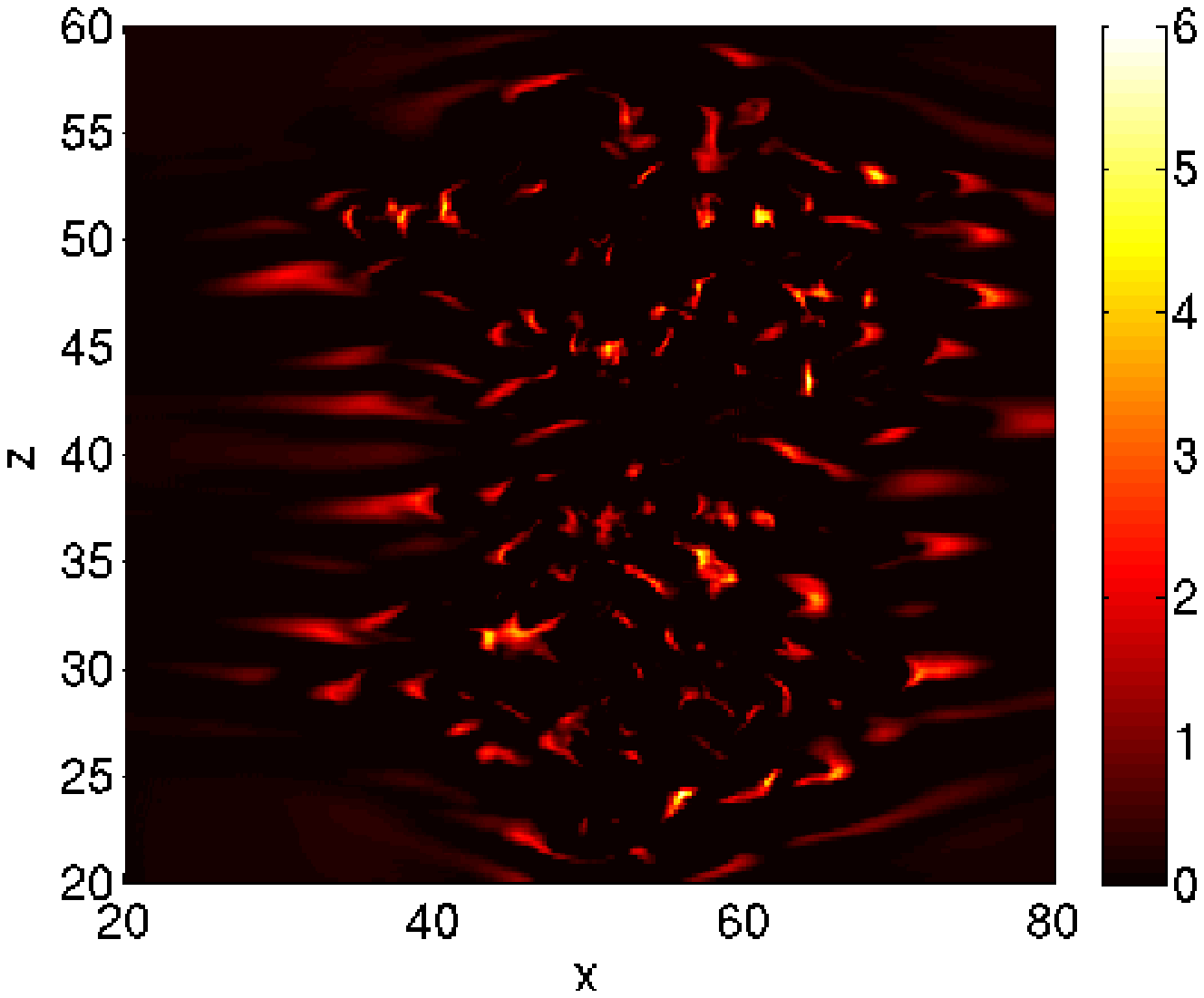}\textbf{(b)}\includegraphics[width=6cm]{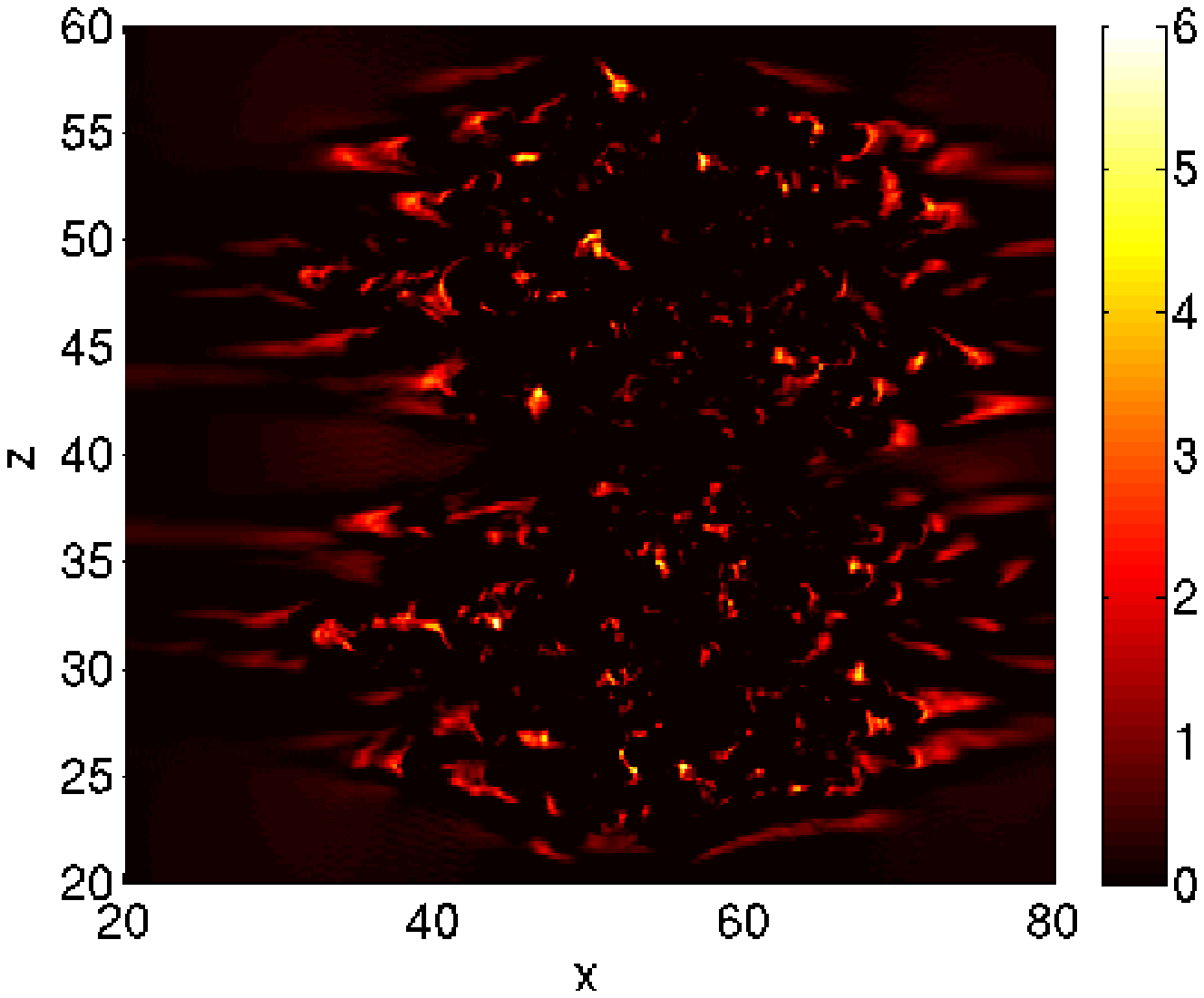}}
\caption{Absolute value of the thresholded spanwise vorticity field in the $y=0$ plane. (a) : at $R=370$, (b) : at $R=500$.}
\label{omegaz}
\end{figure}

  In order to follow the instability, we threshold the spanwise vorticity field in order to keep only $w_z<0$: this yields $\omega_z^{\rm th}$ (Fig.~\ref{omegaz}). Note that the four ``half band'' of the arrow shape spot seem to reveal two cores of production of spanwise vorticity. Besides, we use the result of the laminar-turbulent discrimination to construct a spatial mask $I^{\rm t,i}$ which is $1$ in the turbulent (resp. intermediate) zone and $0$ else where. We then define : $\omega_z^{\rm i,th}=I^{\rm i}\omega_z^{\rm th}$ and $\omega_z^{\rm
t,th}=I^{\rm t}\omega_z^{\rm th}$. This procedure will be discussed in details in the study of the secondary instability in the bands \cite{isp}.

\subsection{Scales of perturbations}

\begin{figure}
\centerline{\includegraphics[height=5cm,clip]{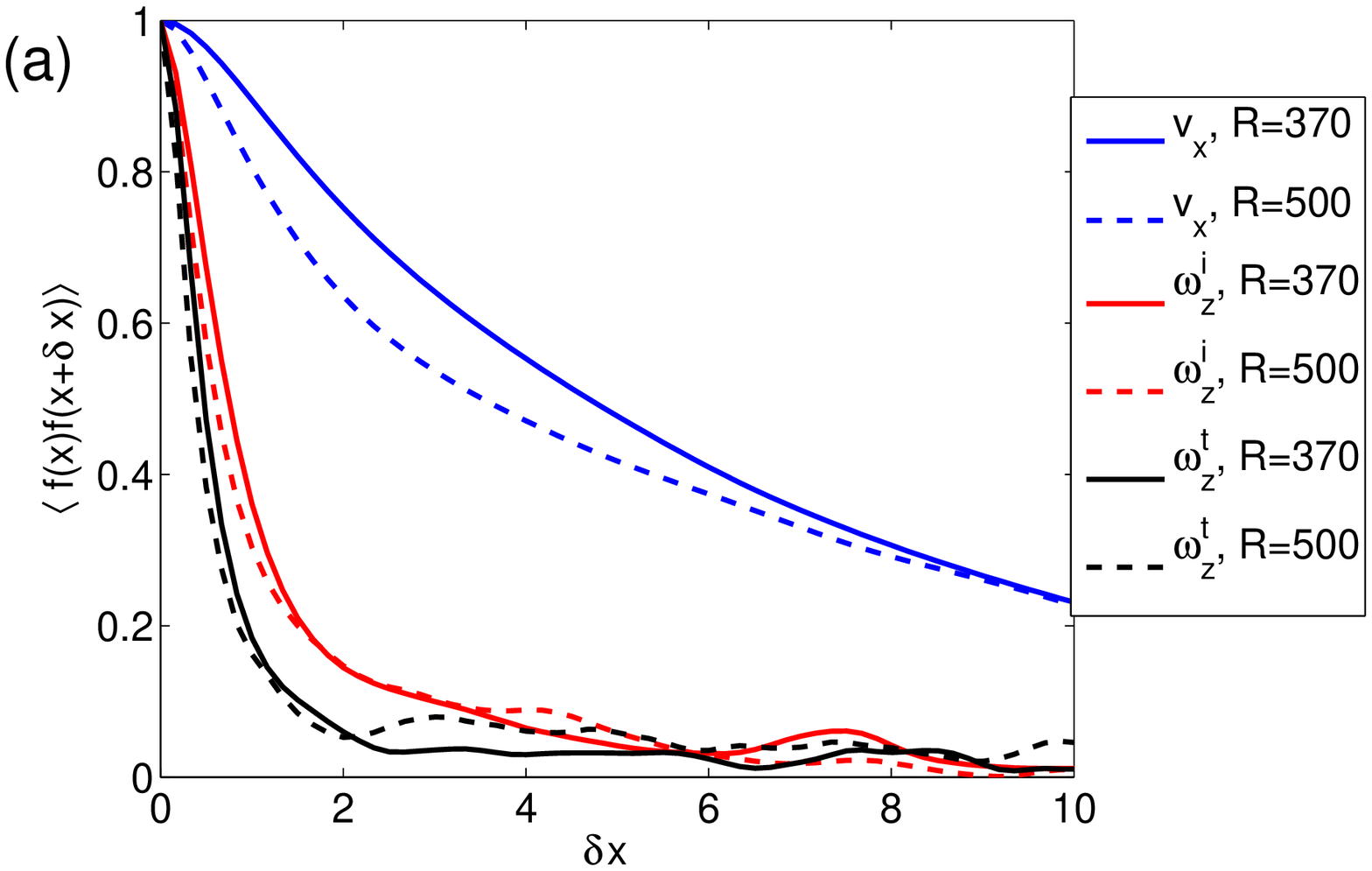}\includegraphics[height=5cm,clip]{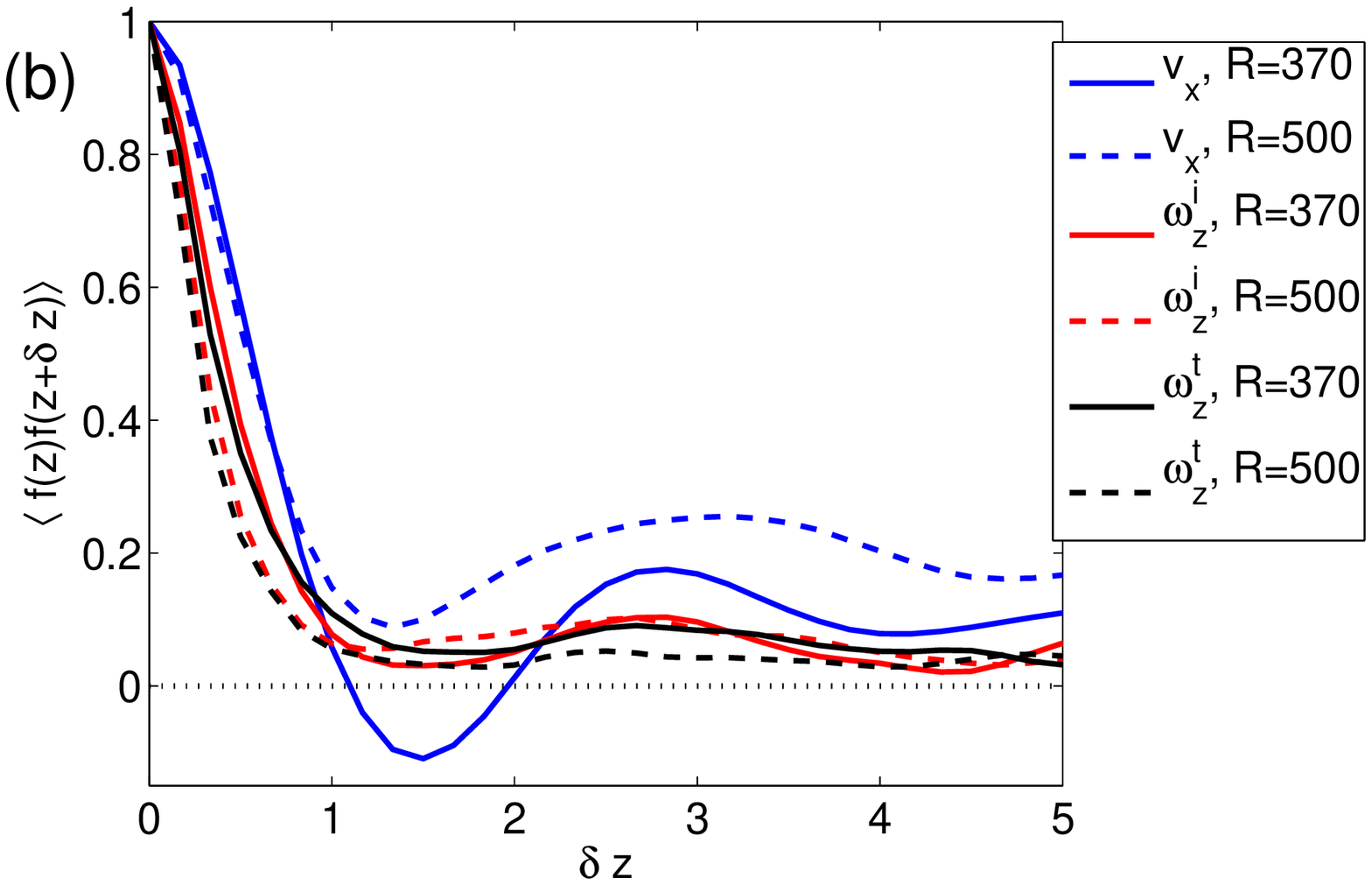}}
\caption{Correlation function of streamwise velocity and thresholded and spatially filtered spanwise vorticity, at $R=370$ and $R=500$. (a): streamwise direction, (b): spanwise direction.}
\label{corspot}
\end{figure}

  We intend to compare the spatial scales of the secondary instability to that of the velocity field, in both intermediate and turbulent region. We therefore compute
the normalised correlation function $\langle f(x,z)f(x+\delta x,z+\delta z)\rangle$ of thresholded and masked spanwise
vorticity $\omega_z^{\rm th,i,t}$ and that of $v_x$. We separate the streamwise and spanwise dependence and present the results in the streamwise ($\delta_z=0$, Fig.~\ref{corspot} (a))
and spanwise ($\delta_x=0$, Fig.~\ref{corspot} (b)) directions. These correlation functions are not time
averaged, unlike what will be done for the bands \cite{isp}.

  The global picture on the scale of perturbations in spots
 appeared in the visualisations of $\omega_z^{\rm th}$ (Fig.~\ref{omegaz}). The spatial scales are characterised by the correlation length $\zeta$, the
inverse of the slope of the correlation function at $\delta
x,\delta z=0$. The streamwise correlation length of the thresholded vorticity field, of order $1$, is smaller in the
turbulent zone ($\zeta\simeq 1$) than in the intermediate one ($\zeta\simeq 1.5$) (Fig.~\ref{corspot}
(a)). The streamwise
velocity $v_x$ correlation function measures the streamwise coherence
length of the velocity streaks. It is one
order of magnitude larger than that of the perturbations. These sizes do not depend much on the
Reynolds number.
This illustrates the separation of scales between the two
fields \cite{PRL,isp}. The perturbations to the streaks baseflow have
the same spanwise size in the turbulent or the intermediate zones
(Fig.~\ref{corspot} (b)). It is of order $h$. The spanwise
correlation length is larger at $R=370$ than it is at $R=500$.
Spanwise quasi-periodicity of the streamwise velocity field leaves a
trace in the correlation function: there is modulation, and the
envelope decreases slowly. Whereas the envelope decreases fast for
the spanwise vorticity $\omega_z^{{\rm i},{\rm t},{\rm th}}$. The trace
of periodicity found in the spanwise vorticity field is a
consequence of the periodicity of the background on which it
develops, as it is the case in the bands \cite{isp}.

\subsection{Advection of perturbations}

  In the band case, the spanwise
advection velocity $c_z(x,z)$ of perturbations is measured \emph{via} a PIV type correlation procedure of the thresholded vorticity field $\omega_z^{\rm th}$.
For that matter, each $x-z$ plane is divided in squares of size $2\times 2$ labeled by the position of its ``lower left'' corner $x_0,z_0$. For each square and at each time $t$, we compute the correlation function $C_{x_0,z_0,y,t}(\Delta z,\delta t)$ between the thresholded vorticity at time $t$, and that at time $t+\delta t$ in order to determine in which direction and by how much the perturbation present here has moved. One has:
\begin{equation}
C_{x_0,z_0,y,t}(\Delta z,\delta t)=\int_{x_0,z_0}^{x_0+2,z_0+2}{\rm d}x{\rm d}z\, \tilde{\omega}(x,y,z,t)\tilde{\omega}(x,y,z+\Delta z,t+\delta t)\,.
\end{equation}
The field $\tilde{\omega}$ denominates $\left(\omega_z^{\rm th}-\langle
\omega_z^{\rm th} \rangle\right)/(\langle(\omega_z^{\rm th}-\langle
\omega_z^{\rm th} \rangle)^2 \rangle)^\frac12$. Here, $\langle .\rangle$ is the average over the relevant $2$ by $2$ square. At fixed $\delta t$, we determine the value $\Delta z$ for which the correlation function is maximum. We then define: $c_z(x,z)=\Delta z/\delta t$.

 This procedure will be studied in details in the case of the band \cite{isp}. In our case no time
average can be performed, due to the expansion of the spot. No
average along the diagonal direction can be performed either, since there
is no favored direction. The band case reveals that the advection
velocity are consistent over the wall normal direction. Therefore a
$y$ average is performed to smooth the results. Only data on the
spanwise advection velocity is obtained in the spot case: the band case shows that this field
has a higher signal over noise ratio and requires time averages.

\begin{figure}
\centerline{\includegraphics[width=6cm,clip]{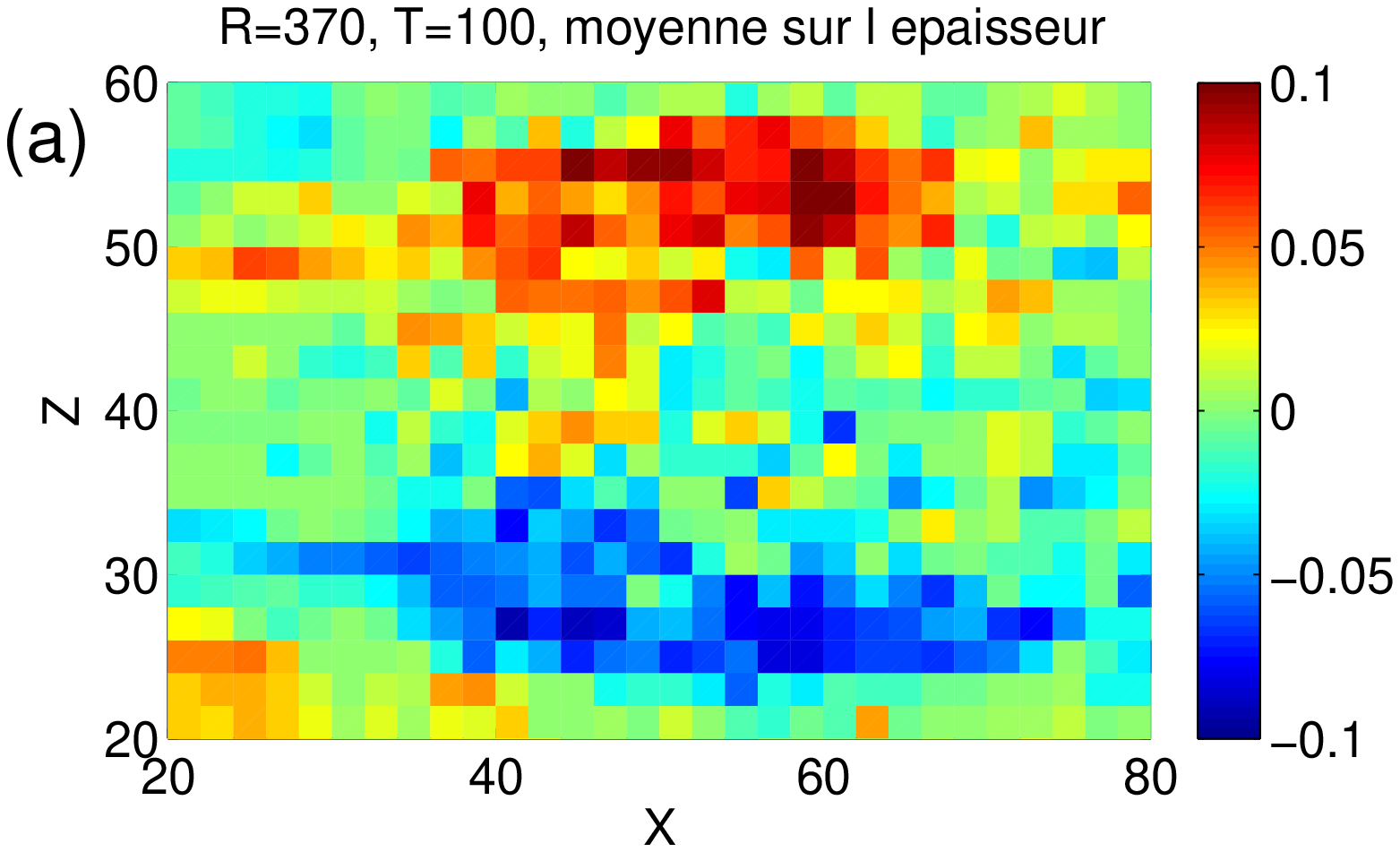}
\includegraphics[width=6cm,clip]{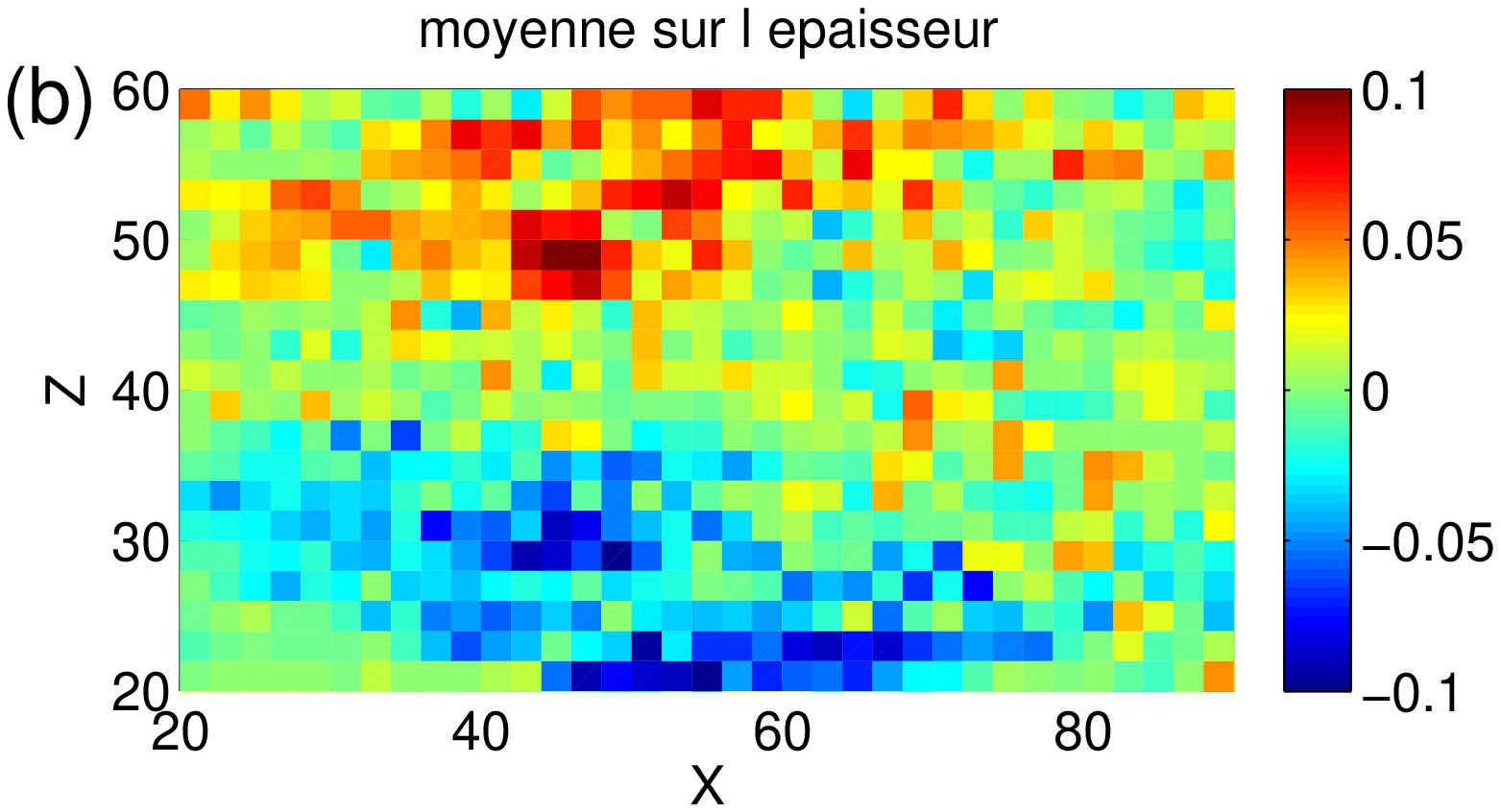}}
\caption{Field of advection velocity, in the case of spots, averaged
over $y$ (a): $R=370$ (b): $R=500$. The figures are zoomed on the
spots} \label{advspdsp}
\end{figure}

  The procedure is applied to the two examples of early stages of spots
considered (figure~\ref{advspdsp} (a) and (b)). No qualitative
difference is found between the $R=370<R_{\rm t}$ case and the
$R=500>R_{\rm t}$ one. The order of magnitude of the spanwise advection velocity is between $-0.05$ and $0.05$. It is in good agreement
with the order of magnitude of the average velocity field as to the direction, different in the $+$ and $-$ parts of the spot (Fig.~\ref{average_spot} (a)), as well as to the value of the advection velocity
(Fig.~\ref{average_spot} (b)). Therefore part of the
quadrupolar flow picture is found in this result. perturbations that
arise in any part of the spot are advected toward the sides.

  The study of the bands shows that an average over the diagonal is necessary to show the exact correspondence between the advection velocity of perturbations and the wall normal average of the velocity field. However, given the agreement in term of order of magnitude as well as direction between the spanwise large scale flow and the spanwise advection velocity, one can expect a similar microscopic behaviour in spots as in
bands in term of advection of perturbations by the large scale flow \cite{PRL,isp}. The geometry of the large scale flow is the only difference. The combined
effect of the streamwise and spanwise velocity is likely to advect
perturbations appearing in the intermediate area along the streamwise
limit spot toward the spanwise limits. The perturbations appearing in the
turbulent area are not likely be advected.

\section{Final remarks \label{D}}

  The growth and reorganisation of the turbulent spot was studied in plane Couette flow.
The quantities used to monitor this phenomenon were validated in the
study of the turbulent bands \cite{RM}. It shows that for $R<R_{\rm
t}$ and for $R>R_{\rm t}$, the spot grows at constant speed: the likelihood of spanwise
expansion is much larger than that of retraction at that Reynolds
number \cite{Yms}. At $R>R_{\rm t}$, the whole domain is invaded by
turbulence. At $R<R_{\rm t}$, a turbulent band forms after a
reorganisation transient. The growth speed of the domain is not monotonous in that stage. It is perturbed by the reorganisation, mostly due to the symmetry breaking, and elimination of defects. The time evolution of the order parameters is stochastic and depend strongly on the initial conditions. It is expected that the stochastic duration of this
reorganisation regime increases with the number of original metastable defects, whose number increases with the domain size. The intrinsic turbulent
noise leads the system toward an optimum, given the boundary
conditions.

  The secondary Kelvin--Helmholtz type instability reported in the bands \cite{etc} and in pipe flow \cite{SK,DWK} is found in the spots as well. The
streamwise and spanwise lengthscales are unchanged by the spot or
band configuration. The perturbations are advected in a direction and with a velocity very close to that of the large
scale flow around the spot. The difference lies in the large scale organisation of
turbulence. The same findings are expected for spot growth in large
domains \cite{DSH}, where they can take various shapes. Moreover, no
qualitative difference is found between the $R<R_{\rm t}$ and the
$R>R_{\rm t}$ cases. The lengthscales, although smaller in spanwise
direction, are of the same order of magnitude. The so-called secondary instability can
be invoked in cycles of the same kind as those proposed for
Poiseuille pipe flow \cite{SK}. Vorticity is advected toward the
spanwise interface of the spot can contribute to the excitation of
the interface.

  Together with the large scale flow, the rolls can stimulate
the stochastic mechanism of spanwise growth \cite{Yms}. However the precise mechanical mechanism of the creation of streaks and vorticies on the side of the spot is unknown. The interaction of the spanwise vorticity is a matter of future investigations. Linear modeling \cite{etc,isp} explains the onset of the instability at the local level and the effect of the advection, refinements taking into account the time evolution of the baseflow can be used in the case of the spot \cite{ACC}. Non-linear models can shed light on the interaction between this instability and the other processes found in the flow, to improve the understanding of the spot growth process. No qualitative differences between the $R=370<R_{\rm t}$ and the $R=500>R_{\rm t}$. This fact
suggests that this instability \emph{alone} has no effect on the
disappearance of the bands as $R$ grows toward $R_{\rm t}$. Other effects are at play: the appearance of laminar troughs for $R$
below $R_{\rm t}$ in an uniformly turbulent flow has been described in DNS \cite{M11,holes} as well as in models \cite{B}.

\section*{acknowledgments}
The author acknowledges discussions with Y. Duguet and P. Manneville and helpful comments from the reviewers.

\end{document}